\documentclass[12pt]{article}
\def\citep#1{\cite{#1}}

\newif\ifpreprint
\preprintfalse

\usepackage{fullpage}

\usepackage[utf8]{inputenc} 
\usepackage[T1]{fontenc}    

\usepackage{amsfonts}       
\usepackage{amssymb}
\usepackage{amsmath}
\usepackage{euscript}
\usepackage{stmaryrd} 
\usepackage{color}
\usepackage{cases}
\usepackage[colors]{optsys}
\usepackage{causal}
\usepackage{snippets}

\newcommand{\SnippetChains}{
  \sep
  \begin{sentence}
    \begin{input}
      ~~\PY{k+kn}{Definition}~\PY{n+nf}{ChainsB}~\PY{o}{:=}~\PY{o}{@}\PY{n}{Chains}~\PY{n}{SType}~\PY{n}{leSet1}\PY{o}{.}~
    \end{input}
  \end{sentence}
}
\newcommand{\SnippetMono}{
  \sep
  \begin{sentence}
    \begin{input}
      ~~\PY{k+kn}{Definition}~\PY{n+nf}{Mono}~\PY{o}{:=}~~\PY{n}{Eb}\PY{o}{.+}~\PY{o}{\(\cup\)}~\PY{n}{Er}\PY{o}{.+.}
    \end{input}
  \end{sentence}
}
\newcommand{\SnippetSType}{
  \sep
  \begin{sentence}
    \begin{input}
      ~~\PY{k+kn}{Definition}~\PY{n+nf}{SType}~\PY{o}{:=}~\PY{o}{\PYZob{}}\PY{n}{S}~\PY{o}{|}~\PY{n}{RelIndep}~\PY{n}{Mono}~\PY{n}{S}\PY{o}{\(\,\wedge\,\)}\PY{n}{S}\PY{o}{:\PYZsh{}(}\PY{n}{Er}\PY{o}{.+)}~\PY{o}{\(\subset\)}~\PY{n}{Mono}\PY{o}{\PYZsh{}}\PY{n}{S}\PY{o}{\(\,\wedge\,\)}\PY{n}{S}~\PY{o}{!=}~\PY{n}{set0}\PY{o}{\PYZcb{}.}
    \end{input}
  \end{sentence}
}
\newcommand{\SnippetScal}{
  \sep
  \begin{sentence}
    \begin{input}
      ~~\PY{k+kn}{Definition}~\PY{n+nf}{Scal}~\PY{o}{:=}~\PY{o}{[}\PY{n+nb}{set}~\PY{n}{S}\PY{o}{|}~\PY{n}{RelIndep}~\PY{n}{Mono}~\PY{n}{S}\PY{o}{\(\,\wedge\,\)}\PY{n}{S}\PY{o}{:\PYZsh{}(}\PY{n}{Er}\PY{o}{.+)}~\PY{o}{\(\subset\)}~\PY{n}{Mono}\PY{o}{\PYZsh{}}\PY{n}{S}\PY{o}{\(\,\wedge\,\)}\PY{n}{S}~\PY{o}{!=}~\PY{n}{set0}~\PY{o}{].}
    \end{input}
  \end{sentence}
}
\newcommand{\SnippetScalnotempty}{
  \sep
  \begin{sentence}
    \begin{input}
      ~~\PY{k+kn}{Lemma}~\PY{n+nf}{Scal\PYZus{}not\PYZus{}empty}~\PY{o}{(}\PY{n+nv}{A1}\PY{o}{:}~\PY{n}{NotEmpty}~\PY{n}{T}\PY{o}{)}~\PY{o}{(}\PY{n+nv}{A2}\PY{o}{:}~\PY{o}{\(\neg\)}~\PY{o}{(}\PY{n}{iic}~\PY{o}{(}\PY{n}{Asym}~\PY{n}{Er}\PY{o}{.+))):}\nl
      ~~~~\PY{k+kr}{\(\exists\)}~\PY{n+nv}{v}\PY{o}{,}~\PY{n}{Scal}~\PY{o}{[}\PY{n+nb}{set}~\PY{n}{v}\PY{o}{].}
    \end{input}
  \end{sentence}
}
\newcommand{\SnippetleSetone}{
  \sep
  \begin{sentence}
    \begin{input}
      ~~\PY{k+kn}{Definition}~\PY{n+nf}{leSet1}~\PY{o}{(}\PY{n+nv}{AB}\PY{o}{:}~\PY{n}{SType}\PY{o}{*}\PY{n}{SType}\PY{o}{)}~\PY{o}{:=}\nl
      ~~~~\PY{n}{leSet}~\PY{o}{(}\PY{n}{Asym}~\PY{n}{Eb}\PY{o}{.+)}~\PY{o}{((}\PY{n}{sval}~\PY{n}{AB}\PY{o}{.}\PY{l+m+mi}{1}\PY{o}{),}~\PY{o}{(}\PY{n}{sval}~\PY{n}{AB}\PY{o}{.}\PY{l+m+mi}{2}\PY{o}{)).}\nl
    \end{input}
  \end{sentence}
  \sep
  \begin{sentence}
    \begin{input}
      ~~\PY{k+kn}{Notation}~\PY{l+s+s2}{\PYZdq{}A~[\PYZlt{}=]~B\PYZdq{}}~\PY{o}{:=}~\PY{o}{(}\PY{n}{leSet1}~\PY{o}{(}\PY{n}{A}\PY{o}{,}\PY{n}{B}\PY{o}{)).}
    \end{input}
  \end{sentence}
}
\newcommand{\Snippetleset}{
  \sep
  \begin{sentence}
    \begin{input}
      ~~\PY{k+kn}{Definition}~\PY{n+nf}{leSet}~\PY{n+nv}{T}~\PY{n+nv}{R}\PY{o}{:}~\PY{n}{relation}~\PY{o}{(}\PY{n+nb}{set}~\PY{n}{T}\PY{o}{)}~\PY{o}{:=}~\nl
      ~~\PY{o}{[}\PY{n+nb}{set}~\PY{n}{AB}~\PY{o}{|}\PY{k+kr}{\(\forall\)}~\PY{o}{(}\PY{n+nv}{a}\PY{o}{:}\PY{n}{T}\PY{o}{),}~\PY{o}{(}\PY{n}{a}~\(\in\)~\PY{n}{AB}\PY{o}{.}\PY{l+m+mi}{1}\PY{o}{)}~\PY{o}{\(\rightarrow\)}~\PY{k+kr}{\(\exists\)}~\PY{n+nv}{b}\PY{o}{,}~\PY{n}{b}~\(\in\)~\PY{n}{AB}\PY{o}{.}\PY{l+m+mi}{2}~\PY{o}{\(\,\wedge\,\)}~\PY{o}{(}~\PY{n}{a}~\PY{o}{=}~\PY{n}{b}~\PY{o}{\(\,\vee\,\)}~\PY{n}{R}~\PY{o}{(}\PY{n}{a}\PY{o}{,}\PY{n}{b}\PY{o}{))}~\PY{o}{].}\nl
    \end{input}
  \end{sentence}
  \sep
  \begin{txt}
    \nl
  \end{txt}
  \sep
  \begin{sentence}
    \begin{input}
      ~~\PY{k+kn}{Notation}~\PY{l+s+s2}{\PYZdq{}A~[\PYZlt{}=~R~]~B\PYZdq{}}~\PY{o}{:=}~\PY{o}{(}\PY{n}{leSet}~\PY{n}{R}~\PY{o}{(}\PY{n}{A}\PY{o}{,}\PY{n}{B}\PY{o}{)).}
    \end{input}
  \end{sentence}
}
\newcommand{\SnippetlesetI}{
  \sep
  \begin{sentence}
    \begin{input}
      ~~\PY{k+kn}{Lemma}~\PY{n+nf}{Ile}~\PY{n+nv}{R}~\PY{n+nv}{A}~\PY{n+nv}{B}\PY{o}{:}~\PY{n}{A}~\PY{o}{\(\subset\)}~\PY{n}{B}~\PY{o}{\(\rightarrow\)}~\PY{n}{A}~\PY{o}{[\PYZlt{}=}~\PY{n}{R}\PY{o}{]}~\PY{n}{B}\PY{o}{.}
    \end{input}
  \end{sentence}
}
\newcommand{\SnippetElt}{
  \sep
  \begin{sentence}
    \begin{input}
      ~~\PY{k+kn}{Definition}~\PY{n+nf}{Elt}~\PY{o}{(}\PY{n+nv}{C}\PY{o}{:}~\PY{n+nb}{set}~\PY{n}{SType}\PY{o}{)}~\PY{o}{:=}~\PY{o}{\PYZob{}}\PY{n}{x}~\PY{o}{:}~\PY{n}{T}~\PY{o}{|}\PY{k+kr}{\(\exists\)}~\PY{o}{(}\PY{n+nv}{S}\PY{o}{:}~\PY{n}{SType}\PY{o}{),}~\PY{n}{S}~\(\in\)~\PY{n}{C}~\PY{o}{\(\,\wedge\,\)}~\PY{n}{x}~\(\in\)~\PY{o}{(}\PY{n}{sval}~\PY{n}{S}\PY{o}{)\PYZcb{}.}
    \end{input}
  \end{sentence}
}
\newcommand{\SnippetSinf}{
  \sep
  \begin{sentence}
    \begin{input}
      ~~\PY{k+kn}{Definition}~\PY{n+nf}{Sinf}~\PY{o}{:=}~\nl
      ~~~~\PY{o}{[}~\PY{n+nb}{set}~\PY{n}{v}\PY{o}{:}~\PY{n}{T}~\PY{o}{|}~\nl
      ~~~~~~\PY{k+kr}{\(\exists\)}~\PY{n+nv}{S}\PY{o}{,}~\PY{o}{(}\PY{n}{S}~\(\in\)~\PY{n}{C}\PY{o}{)}~\PY{o}{\(\,\wedge\,\)}~\PY{o}{(}\PY{n}{v}~\(\in\)~\PY{o}{(}\PY{n}{sval}~\PY{n}{S}\PY{o}{))}~\PY{o}{\(\,\wedge\,\)}\nl
      ~~~~~~~~~~~~~\PY{o}{(}\PY{k+kr}{\(\forall\)}~\PY{n+nv}{T}\PY{o}{,}~\PY{n}{T}~\(\in\)~\PY{n}{C}~\PY{o}{\(\rightarrow\)}~\PY{n}{S}~\PY{o}{[\PYZlt{}=]}~\PY{n}{T}~\PY{o}{\(\rightarrow\)}~\PY{n}{v}~\(\in\)~\PY{o}{(}\PY{n}{sval}~\PY{n}{T}\PY{o}{))].}
    \end{input}
  \end{sentence}
}
\newcommand{\SnippetRC}{
  \sep
  \begin{sentence}
    \begin{input}
      ~~\PY{k+kn}{Definition}~\PY{n+nf}{RC}\PY{o}{:=}~\PY{o}{[}\PY{n+nb}{set}~\PY{n}{xy}\PY{o}{:}~\PY{o}{(}\PY{n}{Elt}~\PY{n}{C}\PY{o}{)*(}\PY{n}{Elt}~\PY{n}{C}\PY{o}{)}~\PY{o}{|}\nl
      ~~~~~~~~~~~~~~~~~~~~\PY{o}{((}\PY{n}{sval}~\PY{n}{xy}\PY{o}{.}\PY{l+m+mi}{1}\PY{o}{)}~\(\in\)~\PY{n}{Sinf}~\PY{o}{\(\,\wedge\,\)}~\PY{n}{xy}\PY{o}{.}\PY{l+m+mi}{2}~\PY{o}{=}~\PY{n}{xy}\PY{o}{.}\PY{l+m+mi}{1}\PY{o}{)}\nl
      ~~~~~~~~~~~~~~~~~~~~\PY{o}{\(\,\vee\,\)}~\PY{o}{(\PYZti{}}~\PY{o}{((}\PY{n}{sval}~\PY{n}{xy}\PY{o}{.}\PY{l+m+mi}{1}\PY{o}{)}~\(\in\)~\PY{n}{Sinf}\PY{o}{)}~\PY{o}{\(\,\wedge\,\)}\nl
      ~~~~~~~~~~~~~~~~~~~~~~~~~\PY{o}{(}\PY{n}{Asym}~\PY{n}{Eb}\PY{o}{.+)}~\PY{o}{(}\PY{n}{sval}~\PY{n}{xy}\PY{o}{.}\PY{l+m+mi}{1}\PY{o}{,}~\PY{n}{sval}~\PY{n}{xy}\PY{o}{.}\PY{l+m+mi}{2}\PY{o}{))].}
    \end{input}
  \end{sentence}
}
\newcommand{\Snippetlesetporder}{
  \sep
  \begin{sentence}
    \begin{input}
      ~~\PY{k+kn}{Lemma}~\PY{n+nf}{leSet2\PYZus{}porder}~\PY{n+nv}{R}\PY{o}{:}~\nl
      ~~~~\PY{n}{sporder}~\PY{n}{R}~\PY{o}{\(\rightarrow\)}~\nl
      ~~~~\PY{o}{@}\PY{n}{porder}~\PY{o}{\PYZob{}}\PY{n}{S}\PY{o}{:}~\PY{n+nb}{set}~\PY{n}{T}\PY{o}{|}~\PY{n}{RelIndep}~\PY{n}{R}~\PY{n}{S}\PY{o}{\PYZcb{}}~\PY{o}{[}\PY{n+nb}{set}~\PY{n}{AB}~\PY{o}{|}~\PY{o}{(}\PY{n}{sval}~\PY{n}{AB}\PY{o}{.}\PY{l+m+mi}{1}\PY{o}{)}~\PY{o}{[\PYZlt{}=}~\PY{n}{R}\PY{o}{]}~\PY{o}{(}\PY{n}{sval}~\PY{n}{AB}\PY{o}{.}\PY{l+m+mi}{2}\PY{o}{)].}
    \end{input}
  \end{sentence}
}
\newcommand{\SnippettotalRC}{
  \sep
  \begin{sentence}
    \begin{input}
      ~~\PY{k+kn}{Lemma}~\PY{n+nf}{total\PYZus{}RC}\PY{o}{:}~\PY{n}{total\PYZus{}rel}~\PY{n}{RC}\PY{o}{.}~
    \end{input}
  \end{sentence}
}
\newcommand{\SnippettotalRCPTr}{
  \sep
  \begin{sentence}
    \begin{input}
      ~~\PY{k+kn}{Lemma}~\PY{n+nf}{total\PYZus{}RC\PYZus{}P3}\PY{o}{:}\nl
      ~~~~\PY{o}{\(\neg\)}~\PY{o}{(}\PY{n}{iic}~\PY{o}{(}\PY{n}{Asym}~\PY{n}{Eb}\PY{o}{.+))}~\PY{o}{\(\rightarrow\)}\nl
      ~~~~\PY{k+kr}{\(\forall\)}~\PY{n+nv}{s}\PY{o}{,}~\PY{k+kr}{\(\exists\)}~\PY{n+nv}{f}\PY{o}{,}~\PY{n}{f}~\PY{l+m+mi}{0}\PY{o}{=}\PY{n}{s}~\PY{o}{\(\,\wedge\,\)}~\PY{o}{(}\PY{k+kr}{\(\exists\)}~\PY{n+nv}{n}\PY{o}{,}~\PY{o}{(}\PY{n}{sval}~\PY{o}{(}\PY{n}{f}~\PY{n}{n}\PY{o}{))}~\(\in\)~\PY{n}{Sinf}~\PY{o}{\(\,\wedge\,\)}~\PY{n}{RC}~\PY{o}{((}\PY{n}{f}~\PY{l+m+mi}{0}\PY{o}{),}~\PY{o}{(}\PY{n}{f}~\PY{n}{n}\PY{o}{))).}
    \end{input}
  \end{sentence}
}
\newcommand{\SnippetChooseRCCi}{
  \sep
  \begin{sentence}
    \begin{input}
      ~~\PY{k+kn}{Lemma}~\PY{n+nf}{ChooseRC5}\PY{o}{:\PYZti{}}~\PY{o}{(}\PY{n}{iic}~\PY{o}{(}\PY{n}{Asym}~\PY{n}{Eb}\PY{o}{.+))}\nl
      ~~~~~~~~\PY{o}{\(\rightarrow\)}~\PY{k+kr}{\(\forall\)}~\PY{o}{(}\PY{n+nv}{s}\PY{o}{:}~\PY{n}{Elt}~\PY{n}{C}\PY{o}{),}~\PY{o}{(}\PY{n}{sval}~\PY{n}{s}~\(\in\)~\PY{n}{Sinf}\PY{o}{)}~\PY{o}{\(\,\vee\,\)}~\nl
      ~~~~~~~~~~~~~~~~~~~~\PY{k+kr}{\(\exists\)}~\PY{o}{(}\PY{n+nv}{s\PYZsq{}}\PY{o}{:}\PY{n}{T}\PY{o}{),}~\PY{o}{(}\PY{n}{s\PYZsq{}}~\(\in\)~\PY{n}{Sinf}\PY{o}{)}~\PY{o}{\(\,\wedge\,\)}~\PY{o}{(}\PY{n}{Asym}~\PY{n}{Eb}\PY{o}{.+)}~\PY{o}{(}\PY{n}{sval}~\PY{n}{s}\PY{o}{,}~\PY{n}{s\PYZsq{}}\PY{o}{).}
    \end{input}
  \end{sentence}
}
\newcommand{\SnippetChooseRCSi}{
  \sep
  \begin{sentence}
    \begin{input}
      ~~\PY{k+kn}{Lemma}~\PY{n+nf}{ChooseRC6}\PY{o}{:\PYZti{}}~\PY{o}{(}\PY{n}{iic}~\PY{o}{(}\PY{n}{Asym}~\PY{n}{Eb}\PY{o}{.+))}\nl
      ~~~~~~~~\PY{o}{\(\rightarrow\)}~\PY{k+kr}{\(\forall\)}~\PY{o}{(}\PY{n+nv}{S}\PY{o}{:}~\PY{n}{SType}\PY{o}{),}~\PY{o}{(}\PY{n}{S}~\(\in\)~\PY{n}{C}\PY{o}{)}~\PY{o}{\(\rightarrow\)}~\PY{o}{(}\PY{n}{sval}~\PY{n}{S}\PY{o}{)}~\PY{o}{[\PYZlt{}=}~\PY{o}{(}\PY{n}{Asym}~\PY{n}{Eb}\PY{o}{.+)]}~\PY{n}{Sinf}\PY{o}{.}
    \end{input}
  \end{sentence}
}
\newcommand{\SnippetSinfScal}{
  \sep
  \begin{sentence}
    \begin{input}
      ~~\PY{k+kn}{Lemma}~\PY{n+nf}{Sinf\PYZus{}Scal}~\PY{o}{(}\PY{n+nv}{A3}\PY{o}{:}~\PY{o}{\(\neg\)}~\PY{n}{iic}~\PY{o}{(}\PY{n}{Asym}~\PY{n}{Eb}\PY{o}{.+)):}~\PY{o}{(}\PY{n}{Sinf}~\PY{n}{C}\PY{o}{)}~\(\in\)~\PY{n}{Scal}\PY{o}{.}~
    \end{input}
  \end{sentence}
}
\newcommand{\SnippetSinfScalP}{
  \sep
  \begin{sentence}
    \begin{input}
      ~~\PY{k+kn}{Lemma}~\PY{n+nf}{Sinf\PYZus{}ScalP}~\PY{o}{(}\PY{n+nv}{A3}\PY{o}{:}~\PY{o}{\(\neg\)}~\PY{n}{iic}~\PY{o}{(}\PY{n}{Asym}~\PY{n}{Eb}\PY{o}{.+)):}~\PY{o}{(}\PY{n}{Sinf}~\PY{n}{C}\PY{o}{):\PYZsh{}(}\PY{n}{Er}\PY{o}{.+)}~\PY{o}{\(\subset\)}~\PY{n}{Mono}\PY{o}{\PYZsh{}(}\PY{n}{Sinf}~\PY{n}{C}\PY{o}{).}
    \end{input}
  \end{sentence}
}

\newcommand{\SnippetSmax}{
  \sep
  \begin{sentence}
    \begin{input}
      ~~\PY{k+kn}{Lemma}~\PY{n+nf}{Smax}~\PY{o}{(}\PY{n+nv}{A1}\PY{o}{:}~\PY{n}{NotEmpty}~\PY{n}{T}\PY{o}{)}~\PY{o}{(}\PY{n+nv}{A2}\PY{o}{:}~\PY{o}{\(\neg\)}~\PY{o}{(}\PY{n}{iic}~\PY{o}{(}\PY{n}{Asym}~\PY{n}{Er}\PY{o}{.+)))}~\PY{o}{(}\PY{n+nv}{A3}\PY{o}{:}~\PY{o}{\(\neg\)}~\PY{o}{(}\PY{n}{iic}~\PY{o}{(}\PY{n}{Asym}~\PY{n}{Eb}\PY{o}{.+))):}\nl
      ~~~~\PY{k+kr}{\(\exists\)}~\PY{n+nv}{Sm}\PY{o}{,}~\PY{n}{IsMaximal}~\PY{n}{Sm}\PY{o}{.}
    \end{input}
  \end{sentence}
}
\newcommand{\SnippetIsMaximal}{
  \sep
  \begin{sentence}
    \begin{input}
      ~~\PY{k+kn}{Definition}~\PY{n+nf}{IsMaximal}~\PY{o}{(}\PY{n+nv}{S}\PY{o}{:}~\PY{n+nb}{set}~\PY{n}{T}\PY{o}{):=}~\nl
      ~~~~\PY{n}{S}~\(\in\)~\PY{n}{Scal}~\PY{o}{\(\,\wedge\,\)}~\PY{k+kr}{\(\forall\)}~\PY{n+nv}{T}\PY{o}{,}~\PY{n}{T}~\(\in\)~\PY{n}{Scal}~\PY{o}{\(\rightarrow\)}~\PY{n}{S}~\PY{o}{[\PYZlt{}=}~\PY{o}{(}\PY{n}{Asym}~\PY{n}{Eb}\PY{o}{.+)]}~\PY{n}{T}~\PY{o}{\(\rightarrow\)}~\PY{n}{T}~\PY{o}{=}~\PY{n}{S}\PY{o}{.}
    \end{input}
  \end{sentence}
}
\newcommand{\SnippetTmI}{
  \sep
  \begin{sentence}
    \begin{input}
      ~~\PY{k+kn}{Lemma}~\PY{n+nf}{TmI}\PY{o}{:}~\PY{k+kr}{\(\forall\)}~\PY{n+nv}{x}\PY{o}{,}~\PY{n}{Tm}~\PY{n}{x}~\PY{o}{\(\subset\)}~\PY{n}{Sm}\PY{o}{.}
    \end{input}
  \end{sentence}
}
\newcommand{\SnippetSbunp}{
  \sep
  \begin{sentence}
    \begin{input}
      ~~\PY{k+kn}{Lemma}~\PY{n+nf}{fact0}\PY{o}{:}~\PY{k+kr}{\(\forall\)}~\PY{n+nv}{x}~\PY{n+nv}{y}\PY{o}{,}~\PY{n}{y}~\(\in\)~\PY{n}{Sm}~\PY{o}{`\PYZbs{}`}~\PY{o}{(}\PY{n}{Tm}~\PY{n}{x}\PY{o}{)}~\PY{o}{\(\rightarrow\)}~\PY{n}{Eb}\PY{o}{.+}~\PY{o}{(}\PY{n}{y}\PY{o}{,}\PY{n}{x}\PY{o}{).}
    \end{input}
  \end{sentence}
}

\newcommand{\Snippetfactnine}{
  \sep
  \begin{sentence}
    \begin{input}
      ~~\PY{k+kn}{Lemma}~\PY{n+nf}{fact9}\PY{o}{:}~\PY{n}{IsMaximal}~\PY{n}{Sm}~\PY{o}{\(\rightarrow\)}~\PY{o}{(}\PY{k+kr}{\(\forall\)}~\PY{n+nv}{x}\PY{o}{,}~\PY{n}{x}~\(\in\)~\PY{n}{Se}~\PY{o}{\(\rightarrow\)}~\PY{n}{SeP}~\PY{n}{x}~\PY{o}{\(\rightarrow\)}~\PY{k+kt}{False}\PY{o}{).}
    \end{input}
  \end{sentence}
}
\newcommand{\Snippetfactten}{
  \sep
  \begin{sentence}
    \begin{input}
      ~~\PY{k+kn}{Lemma}~\PY{n+nf}{fact10}~\PY{o}{(}\PY{n+nv}{A2}\PY{o}{:}~\PY{o}{\(\neg\)}~\PY{o}{(}\PY{n}{iic}~\PY{o}{(}\PY{n}{Asym}~\PY{n}{Er}\PY{o}{.+))):}~\PY{n}{IsMaximal}~\PY{n}{Sm}~\PY{o}{\(\rightarrow\)}~\PY{o}{\(\neg\)(}\PY{k+kr}{\(\exists\)}~\PY{n+nv}{x}\PY{o}{,}~\PY{n}{x}~\(\in\)~\PY{n}{Se}\PY{o}{).}
    \end{input}
  \end{sentence}
}
\newcommand{\Snippetfacteleven}{
  \sep
  \begin{sentence}
    \begin{input}
      ~~\PY{k+kn}{Lemma}~\PY{n+nf}{fact11}\PY{o}{:}~~\PY{o}{\(\neg\)(}\PY{k+kr}{\(\exists\)}~\PY{n+nv}{x}\PY{o}{,}~\PY{n}{x}~\(\in\)~\PY{n}{Se}\PY{o}{)}~\PY{o}{\(\rightarrow\)}~\PY{o}{(}\PY{k+kr}{\(\forall\)}~\PY{n+nv}{x}\PY{o}{,}~\PY{o}{\(\neg\)}~\PY{o}{(}\PY{n}{x}\(\in\)~\PY{n}{Sm}\PY{o}{)}~\PY{o}{\(\rightarrow\)}~\PY{o}{(}\PY{n}{x}~\(\in\)~\PY{n}{Mono}\PY{o}{\PYZsh{}}\PY{n}{Sm}\PY{o}{)).}
    \end{input}
  \end{sentence}
}
\newcommand{\SnippetSx}{
  \sep
  \begin{sentence}
    \begin{input}
      ~~\PY{k+kn}{Definition}~\PY{n+nf}{Se}\PY{o}{:=}~\PY{o}{[}\PY{n+nb}{set}~\PY{n}{y}~\PY{o}{|}~\PY{o}{\(\neg\)}~\PY{o}{(}\PY{n}{y}~\(\in\)~\PY{n}{Sm}\PY{o}{)}~\PY{o}{\(\,\wedge\,\)}~\PY{o}{\(\neg\)}~\PY{o}{(}\PY{n}{y}~\(\in\)~\PY{n}{Mono}\PY{o}{\PYZsh{}}\PY{n}{Sm}\PY{o}{)].}
    \end{input}
  \end{sentence}
}
\newcommand{\SnippetTm}{
  \sep
  \begin{sentence}
    \begin{input}
      ~~\PY{k+kn}{Definition}~\PY{n+nf}{Tm}~\PY{n+nv}{x}\PY{o}{:=}~\PY{o}{[}\PY{n+nb}{set}~\PY{n}{y}~\PY{o}{|}~\PY{n}{y}~\(\in\)~\PY{n}{Sm}~\PY{o}{\(\,\wedge\,\)}~\PY{o}{\(\neg\)}~\PY{o}{(}\PY{n}{Eb}\PY{o}{.+}~\PY{o}{(}\PY{n}{y}\PY{o}{,}\PY{n}{x}\PY{o}{))].}
    \end{input}
  \end{sentence}
}
\newcommand{\SnippetSxm}{
  \sep
  \begin{sentence}
    \begin{input}
      ~~\PY{k+kn}{Definition}~\PY{n+nf}{SeP}~\PY{n+nv}{x}~\PY{o}{:=}~\PY{k+kr}{\(\forall\)}~\PY{n+nv}{y}\PY{o}{,}~\PY{n}{y}~\(\in\)~\PY{n}{Se}~\PY{o}{\(\rightarrow\)}~\PY{n}{Er}\PY{o}{.+(}\PY{n}{x}\PY{o}{,}\PY{n}{y}\PY{o}{)}~\PY{o}{\(\rightarrow\)}~\PY{n}{Er}\PY{o}{.+(}\PY{n}{y}\PY{o}{,}\PY{n}{x}\PY{o}{).}
    \end{input}
  \end{sentence}
}
\newcommand{\SnippetSxone}{
  \sep
  \begin{sentence}
    \begin{input}
      ~~\PY{k+kn}{Lemma}~\PY{n+nf}{Sx\PYZus{}1}~\PY{o}{(}\PY{n+nv}{A2}\PY{o}{:}~\PY{o}{\(\neg\)}~\PY{o}{(}\PY{n}{iic}~\PY{o}{(}\PY{n}{Asym}~\PY{n}{Er}\PY{o}{.+))):}\nl
      ~~~~\PY{o}{(}\PY{k+kr}{\(\exists\)}~\PY{o}{(}\PY{n+nv}{x}\PY{o}{:}\PY{n}{T}\PY{o}{),}~\PY{o}{(}\PY{n}{x}~\(\in\)~\PY{n}{Se}\PY{o}{))}~\PY{o}{\(\rightarrow\)}~\PY{o}{(}\PY{k+kr}{\(\exists\)}~\PY{o}{(}\PY{n+nv}{x}\PY{o}{:}\PY{n}{T}\PY{o}{),}~\PY{n}{x}~\(\in\)~\PY{n}{Se}~\PY{o}{\(\,\wedge\,\)}~\PY{n}{SeP}~\PY{n}{x}\PY{o}{).}
    \end{input}
  \end{sentence}
}
\newcommand{\Snippetinfasym}{
  \sep
  \begin{sentence}
    \begin{input}
      ~~\PY{k+kn}{Lemma}~\PY{n+nf}{iic\PYZus{}asym\PYZus{}to\PYZus{}iic\PYZus{}inj}\PY{o}{:}~~\PY{o}{(}\PY{n}{iic}~\PY{o}{(}\PY{n}{Asym}~\PY{n}{R}\PY{o}{.+))}~\PY{o}{\(\rightarrow\)}~\PY{o}{(}\PY{n}{iic\PYZus{}inj}~\PY{n}{R}\PY{o}{).}~
    \end{input}
  \end{sentence}
}
\newcommand{\SnippetMainTh}{
  \sep
  \begin{sentence}
    \begin{input}
      ~~\PY{k+kn}{Theorem}~\PY{n+nf}{SSWext}\PY{o}{:}\nl
      ~~~~\PY{o}{(}\PY{k+kr}{\(\exists\)}~\PY{o}{(}\PY{n+nv}{v0}\PY{o}{:}\PY{n}{T}\PY{o}{),}~\PY{o}{(}\PY{n}{v0}~\(\in\)~\PY{n}{setT}\PY{o}{))}~\PY{o}{\(\rightarrow\)}~\PY{o}{\(\neg\)}~\PY{o}{(}\PY{n}{iic}~\PY{o}{(}\PY{n}{Asym}~\PY{n}{Er}\PY{o}{.+))}~\PY{o}{\(\rightarrow\)}~\PY{o}{\(\neg\)}~\PY{o}{(}\PY{n}{iic}~\PY{o}{(}\PY{n}{Asym}~\PY{n}{Eb}\PY{o}{.+))}\nl
      ~~~~\PY{o}{\(\rightarrow\)}~\PY{k+kr}{\(\exists\)}~\PY{n+nv}{Sm}\PY{o}{,}~\PY{n}{RelIndep}~\PY{n}{Mono}~\PY{n}{Sm}~\PY{o}{\(\,\wedge\,\)}~~\PY{n}{Sm}~\PY{o}{!=}~\PY{n}{set0}~\PY{o}{\(\,\wedge\,\)}~\nl
      ~~~~~~~~~~~~\PY{k+kr}{\(\forall\)}~\PY{n+nv}{x}\PY{o}{,}~\PY{o}{\(\neg\)}~\PY{o}{(}\PY{n}{x}\(\in\)~\PY{n}{Sm}\PY{o}{)}~\PY{o}{\(\rightarrow\)}~\PY{o}{(}\PY{n}{x}~\(\in\)~\PY{n}{Mono}\PY{o}{\PYZsh{}}\PY{n}{Sm}\PY{o}{).}~
    \end{input}
  \end{sentence}
}
\newcommand{\SnippetSSWTh}{
  \sep
  \begin{sentence}
    \begin{input}
      ~~\PY{k+kn}{Corollary}~\PY{n+nf}{SSW}\PY{o}{:}\nl
      ~~~\PY{o}{(}\PY{k+kr}{\(\exists\)}~\PY{o}{(}\PY{n+nv}{v0}\PY{o}{:}\PY{n}{T}\PY{o}{),}~\PY{o}{(}\PY{n}{v0}~\(\in\)~\PY{n}{setT}\PY{o}{))}~\PY{o}{\(\rightarrow\)}~\PY{o}{\(\neg\)}~\PY{o}{(}\PY{n}{iic\PYZus{}inj}~\PY{n}{Er}\PY{o}{)}~\PY{o}{\(\rightarrow\)}~\PY{o}{\(\neg\)}~\PY{o}{(}\PY{n}{iic\PYZus{}inj}~\PY{n}{Eb}\PY{o}{)}\nl
      ~~~\PY{o}{\(\rightarrow\)}~\PY{k+kr}{\(\exists\)}~\PY{n+nv}{Sm}\PY{o}{,}~\PY{n}{RelIndep}~\PY{n}{Mono}~\PY{n}{Sm}~\PY{o}{\(\,\wedge\,\)}~~\PY{n}{Sm}~\PY{o}{!=}~\PY{n}{set0}~\PY{o}{\(\,\wedge\,\)}\nl
      ~~~~~~~~~~~~~\PY{k+kr}{\(\forall\)}~\PY{n+nv}{x}\PY{o}{,}~\PY{o}{\(\neg\)}~\PY{o}{(}\PY{n}{x}\(\in\)~\PY{n}{Sm}\PY{o}{)}~\PY{o}{\(\rightarrow\)}~\PY{o}{(}\PY{n}{x}~\(\in\)~\PY{n}{Mono}\PY{o}{\PYZsh{}}\PY{n}{Sm}\PY{o}{).}~
    \end{input}
  \end{sentence}
}
\newcommand{\SnippetMonotopath}{
  \sep
  \begin{sentence}
    \begin{input}
      ~~\PY{k+kn}{Lemma}~\PY{n+nf}{Mono2path}~\PY{n+nv}{x}~\PY{n+nv}{S}\PY{o}{:}~\nl
      ~~~~\PY{n}{x}~\(\in\)~\PY{n}{Mono}\PY{o}{\PYZsh{}}\PY{n}{S}~\PY{o}{\(\rightarrow\)}\nl
      ~~~~\PY{k+kr}{\(\exists\)}~\PY{n+nv}{y}\PY{o}{,}~\PY{n}{y}~\(\in\)~\PY{n}{S}\nl
      ~~~~~~~\PY{o}{\(\,\wedge\,\)}~\PY{k+kr}{\(\exists\)}~\PY{o}{(}\PY{n+nv}{s}\PY{o}{:}~\PY{n}{seq}~\PY{n}{T}\PY{o}{)}~\PY{o}{,}~\PY{o}{\(\neg\)}~\PY{n}{x}~\(\in\)~\PY{n}{s}~\PY{o}{\(\,\wedge\,\)}~\PY{o}{\(\neg\)}~\PY{n}{y}~\(\in\)~\PY{n}{s}~\PY{o}{\(\,\wedge\,\)}~\PY{n}{uniq}~\PY{n}{s}~\nl
      ~~~~~~~~~~~~~~~~~~~~~~~~\PY{o}{\(\,\wedge\,\)}~\PY{o}{(}\PY{n}{allL}~\PY{n}{Eb}~\PY{n}{s}~\PY{n}{x}~\PY{n}{y}~\PY{o}{\(\,\vee\,\)}~\PY{n}{allL}~\PY{n}{Er}~\PY{n}{s}~\PY{n}{x}~\PY{n}{y}\PY{o}{).}
    \end{input}
  \end{sentence}
}
\newcommand{\SnippetAsymdPV}{
  \sep
  \begin{sentence}
    \begin{input}
      ~~\PY{k+kn}{Lemma}~\PY{n+nf}{Asym2P5}\PY{o}{:}~\nl
      ~~~~\PY{o}{(}\PY{n}{iic}~\PY{o}{(}\PY{n}{Asym}~\PY{n}{R}\PY{o}{.+))}~\PY{o}{\(\rightarrow\)}~\PY{k+kr}{\(\exists\)}~\PY{n+nv}{k}\PY{o}{:}~\PY{n}{nat}~\PY{o}{\(\rightarrow\)}~\PY{n}{T}\PY{o}{,}~\PY{k+kr}{\(\exists\)}~\PY{n+nv}{l}\PY{o}{:}~\PY{n}{nat}~\PY{o}{\(\rightarrow\)}~\PY{n}{seq}~\PY{n}{T}\PY{o}{,}\nl
      ~~~~~~\PY{k+kr}{\(\forall\)}~\PY{n+nv}{n}\PY{o}{,}~\PY{n}{allLu}~\PY{n}{R}~\PY{o}{(}\PY{n}{l}~\PY{n}{n}\PY{o}{)}~\PY{o}{(}\PY{n}{k}~\PY{n}{n}\PY{o}{)}~\PY{o}{(}\PY{n}{k}~\PY{n}{n}\PY{o}{.+}\PY{l+m+mi}{1}\PY{o}{)}~\PY{o}{\(\,\wedge\,\)}~\PY{o}{\(\neg\)}~\PY{n}{R}\PY{o}{.+}~\PY{o}{(}\PY{n}{k}~\PY{n}{n}\PY{o}{.+}\PY{l+m+mi}{1}\PY{o}{,}~\PY{n}{k}~\PY{n}{n}\PY{o}{)}\nl
      ~~~~~~~~~~~~~~~~\PY{o}{\(\,\wedge\,\)}~\PY{n}{uniq}~\PY{o}{((}\PY{n}{l}~\PY{n}{n}\PY{o}{)}~\PY{o}{++}~\PY{o}{(}\PY{n}{l}~\PY{n}{n}\PY{o}{.+}\PY{l+m+mi}{1}\PY{o}{)).}
    \end{input}
  \end{sentence}
}

\newcommand{\SnippetRelIndep}{
  \sep
  \begin{sentence}
    \begin{input}
      ~~\PY{k+kn}{Definition}~\PY{n+nf}{RelIndep}~\PY{o}{(}\PY{n+nv}{T}\PY{o}{:}\PY{k+kt}{Type}\PY{o}{)}~\PY{o}{(}\PY{n+nv}{R}\PY{o}{:}~\PY{n}{relation}~\PY{n}{T}\PY{o}{)}~\PY{o}{(}\PY{n+nv}{S}\PY{o}{:}~\PY{n+nb}{set}~\PY{n}{T}\PY{o}{)}~\PY{o}{:=}\nl
      ~~~~\PY{k+kr}{\(\forall\)}~\PY{o}{(}\PY{n+nv}{x}~\PY{n+nv}{y}\PY{o}{:}\PY{n}{T}\PY{o}{),}~~\PY{n}{x}~\(\in\)~\PY{n}{S}~\PY{o}{\(\rightarrow\)}~\PY{n}{y}~\(\in\)~\PY{n}{S}~\PY{o}{\(\rightarrow\)}~\PY{o}{\(\neg\)(}\PY{n}{x}~\PY{o}{=}~\PY{n}{y}\PY{o}{)}~\PY{o}{\(\rightarrow\)}~\PY{o}{\(\neg\)(}~\PY{n}{R}~\PY{o}{(}\PY{n}{x}\PY{o}{,}\PY{n}{y}\PY{o}{)).}
    \end{input}
  \end{sentence}
}

\newcommand{\Snippetasym}{
  \sep
  \begin{sentence}
    \begin{input}
      ~~\PY{k+kn}{Definition}~\PY{n+nf}{Asym}~\PY{n+nv}{R}\PY{o}{:}~\PY{n}{relation}~\PY{n}{T}~\PY{o}{:=}~\PY{o}{[}\PY{n+nb}{set}~\PY{n}{xy}~\PY{o}{|}~\PY{n}{R}~\PY{n}{xy}~\PY{o}{\(\,\wedge\,\)}~\PY{o}{\(\neg\)}~\PY{o}{(}\PY{n}{R}\PY{o}{\PYZca{}\PYZhy{}}\PY{l+m+mi}{1}~\PY{n}{xy}\PY{o}{)].}
    \end{input}
  \end{sentence}
}
\newcommand{\Snippetlefttotal}{
  \sep
  \begin{sentence}
    \begin{input}
      ~~\PY{k+kn}{Definition}~\PY{n+nf}{total\PYZus{}rel}~\PY{o}{(}\PY{n+nv}{R}\PY{o}{:}~\PY{n}{relation}~\PY{n}{T}\PY{o}{)}~\PY{o}{:=}~\PY{k+kr}{\(\forall\)}~\PY{n+nv}{x}\PY{o}{,}~\PY{k+kr}{\(\exists\)}~\PY{n+nv}{y}\PY{o}{,}~\PY{n}{R}~\PY{o}{(}\PY{n}{x}\PY{o}{,}\PY{n}{y}\PY{o}{).}
    \end{input}
  \end{sentence}
}
\newcommand{\Snippetiic}{
  \sep
  \begin{sentence}
    \begin{input}
      ~~\PY{k+kn}{Definition}~\PY{n+nf}{iic}~\PY{n+nv}{R}~\PY{o}{:=}~\PY{k+kr}{\(\exists\)}~\PY{n+nv}{f}\PY{o}{,}~\PY{k+kr}{\(\forall\)}~\PY{n+nv}{n}\PY{o}{,}~\PY{n}{R}~\PY{o}{((}\PY{n}{f}~\PY{n}{n}\PY{o}{),(}\PY{n}{f}~\PY{o}{(}\PY{n}{S}~\PY{n}{n}\PY{o}{))).}~~
    \end{input}
  \end{sentence}
}
\newcommand{\Snippetiicinj}{
  \sep
  \begin{sentence}
    \begin{input}
      ~~\PY{k+kn}{Definition}~\PY{n+nf}{iic\PYZus{}inj}~\PY{n+nv}{R}~\PY{o}{:=}~\PY{k+kr}{\(\exists\)}~\PY{n+nv}{f}\PY{o}{,}~\PY{o}{(}\PY{k+kr}{\(\forall\)}~\PY{n+nv}{n}\PY{o}{,}~\PY{n}{R}~\PY{o}{((}\PY{n}{f}~\PY{n}{n}\PY{o}{),(}\PY{n}{f}~\PY{o}{(}\PY{n}{S}~\PY{n}{n}\PY{o}{))))}~\PY{o}{\(\,\wedge\,\)}~\PY{n}{injective}~\PY{n}{f}\PY{o}{.}
    \end{input}
  \end{sentence}
}
\newcommand{\Snippetchains}{
  \sep
  \begin{sentence}
    \begin{input}
      ~~\PY{k+kn}{Definition}~\PY{n+nf}{Chains}~\PY{o}{(}\PY{n+nv}{R}\PY{o}{:}~\PY{n}{relation}~\PY{n}{T}\PY{o}{)}~\PY{o}{:=}~\nl
      ~~~~\PY{o}{[}\PY{n+nb}{set}~\PY{n}{C}\PY{o}{:}~\PY{n+nb}{set}~\PY{n}{T}\PY{o}{|}~\PY{k+kr}{\(\forall\)}~\PY{o}{(}\PY{n+nv}{c1}~\PY{n+nv}{c2}\PY{o}{:}~\PY{n}{T}\PY{o}{),}~\PY{n}{C}~\PY{n}{c1}~\PY{o}{\(\rightarrow\)}~\PY{n}{C}~\PY{n}{c2}~\PY{o}{\(\rightarrow\)}~~\PY{n}{R}~\PY{o}{(}\PY{n}{c1}\PY{o}{,}\PY{n}{c2}\PY{o}{)}~\PY{o}{\(\,\vee\,\)}~\PY{n}{R}~\PY{o}{(}\PY{n}{c2}\PY{o}{,}\PY{n}{c1}\PY{o}{)].}
    \end{input}
  \end{sentence}
}

\newcommand{\Snippettcpuniq}{
  \sep
  \begin{sentence}
    \begin{input}
      ~~\PY{k+kn}{Lemma}~\PY{n+nf}{TCP\PYZus{}uniq}~\PY{n+nv}{R}~\PY{n+nv}{x}~\PY{n+nv}{y}\PY{o}{:}~\PY{n}{R}\PY{o}{.+}~\PY{o}{(}\PY{n}{x}\PY{o}{,}\PY{n}{y}\PY{o}{)}~\PY{o}{\(\leftrightarrow\)}~\PY{k+kr}{\(\exists\)}~\PY{n+nv}{s}\PY{o}{,}~\PY{o}{\(\neg\)}~\PY{n}{x}~\(\in\)~\PY{n}{s}~\PY{o}{\(\,\wedge\,\)}~\PY{o}{\(\neg\)}~\PY{n}{y}~\(\in\)~\PY{n}{s}~\PY{o}{\(\,\wedge\,\)}~\PY{n}{uniq}~\PY{n}{s}\nl
      ~~~~~~~~~~~~~~~~~~~~~~~~~~~~~~~~~~~~~~~\PY{o}{\(\,\wedge\,\)}~\PY{n}{allL}~\PY{n}{R}~\PY{n}{s}~\PY{n}{x}~\PY{n}{y}\PY{o}{.}~
    \end{input}
  \end{sentence}
}

\newcommand{\Snippetrelation}{
  \sep
  \begin{sentence}
    \begin{input}
      ~~\PY{k+kn}{Definition}~\PY{n+nf}{relation}~\PY{o}{(}\PY{n+nv}{T}\PY{o}{:}~\PY{k+kt}{Type}\PY{o}{)}~\PY{o}{:=}~\PY{n+nb}{set}~\PY{o}{(}\PY{n}{T}~\PY{o}{*}~\PY{n}{T}\PY{o}{).}
    \end{input}
  \end{sentence}
}

\usepackage{framed}
\usepackage{graphicx}
\graphicspath{{FIGURES/}}

\newcommand{\Order}{{\leqslant}}
\newcommand{\RelOrder}[1]{{\Order_{#1}}}
\newcommand{\ROrder}{\RelOrder{\relation}}
\newcommand{\AstOrder}{\RelOrder{\Asym{\np{\rel{E}_b^+}}}}

\newcommand{\Asym}[1]{#1^{\mathsf{as}}}
\newcommand{\rel}[1]{\mathfrak{#1}}
\newcommand{\dgraph}{{D}}
\newlength{\leftbarwidth}
\newlength{\leftbarsep}
\colorlet{leftbarcolor}{red}

\usepackage{amsmath}
\usepackage{fullpage}
\usepackage[numbers]{natbib}
\usepackage{authblk}
\graphicspath{{./},{FIGURES/}}

\title{A formal proof of the Sands-Sauer-Woodrow theorem using the Rocq prover and mathcomp/ssreflect}

\author[1]{Jean-Philippe Chancelier}
\affil[1]{CERMICS, CNRS, ENPC, Institut Polytechnique de Paris, Marne-la-Vallée, France}

\date{\today}

\begin{document}

\maketitle

\begin{abstract}
  We present a formal proof of the Sands-Sauer-Woodrow (SSW) theorem using the
  Rocq proof assistant and the MathComp/SSReflect library.
  The SSW theorem states that in a directed graph whose edges are colored with
  two colors and that contains no monochromatic infinite outward path, there
  exists an independent set~$S$ of vertices such that every vertex outside~$S$
  can reach~$S$ by a monochromatic path.
  We formalize the graph using two binary relations $\mathfrak{E}_b$ and
  $\mathfrak{E}_r$, representing the blue and red edges
  respectively, and we develop a dedicated library for binary relations
  represented as classical sets.  Beyond formalizing the original SSW theorem,
  we establish a strictly stronger version in which the assumption ``no
  monochromatic infinite outward path'' is replaced by the weaker condition that
  the asymmetric parts of the transitive closures of $\mathfrak{E}_b$ and
  $\mathfrak{E}_r$ admit no infinite outward paths.  The original SSW theorem is
  then recovered as a corollary via a lemma showing that an infinite path for
  the asymmetric part of the transitive closure of a relation implies an
  infinite path for the relation.
\end{abstract}


\section{Introduction and Notations}

The objective of this paper is to present a formal proof of the
Sands-Sauer-Woodrow theorem~\cite{Sands-Sauer-Woodrow-1982}, denoted SSW theorem
in the sequel, using the Rocq prover (formerly known as Coq). It is among the most
celebrated theorems about kernels, and is actually a generalization of the
Gale–Shapley ``marriage'' theorem~\cite{Gale-Shapley}. It is formulated as follows.

\emph{Let $\dgraph$ be a directed graph
  \footnote{A directed graph in~\cite{Sands-Sauer-Woodrow-1982} and in this work
    is possibly infinite and possibly with multiple edges.}
  whose edges are colored with two
  colors, such that $\dgraph$ contains no monochromatic infinite outward path.
  Then, there is an independent set $S$ of vertices of $\dgraph$ such that, for every vertex
  $x$ not in $S$, there is a monochromatic path from $x$ to a vertex of $S$.}

A kernel formulation corollary of this theorem in the finite case is the
following: \emph{Let D be a finite digraph whose edges are colored with two
  colors such that the restriction to each color forms a transitive digraph;
  then D admits a kernel}. Kernels form a fundamental topic of the theory of
digraphs. Introduced by von Neumann and Morgenstern in the context of board
games analysis~\cite{vonNeuman-Morgenstern:1947}, they have found applications
in other areas like social choice theory, economy~\cite{Ayumi} and
logic~\cite{Walicki}. They are still subject of many research works. Several
extensions to more than two colors of the SSW theorem remain conjectures and an
active subject of research. As examples, in~\cite{Langlois-Meunier}, the authors revisit
classical results on kernels in digraphs, in~\cite{Galeana-Rojas} the authors extend the 
SSW theorem to $({\cal A},{\cal B})$-kernels and in~\cite{Bousquet-Lochet-Tomasse}, the
authors provide a significant breakthrough regarding tournaments (a type of
directed graph where every pair of distinct vertices has exactly one directed
edge).


We formalize the SSW theorem by employing binary relations.
Specifically, without loss of generality, we model the directed graph $\dgraph$ using
two relations, $\mathfrak{E}_{b}$ and $\mathfrak{E}_{r}$, which represent the blue and red edges,
respectively, on the set of vertices. By utilizing these relations,
the formal proof adheres to the logical steps established in the original
publication~\cite{Sands-Sauer-Woodrow-1982}.

Furthermore, we show that the SSW theorem can be proved under a broader
condition by replacing the assumption of ``no monochromatic infinite outward
path'' with the following weaker assumption: there is no infinite outward paths
in the two graphs defined respectively by the asymmetric part of the transitive
closure of the relation $\mathfrak{E}_{b}$ and of the relation $\mathfrak{E}_{b}$.

The formal verification that these new assumptions are weaker than the original
ones constitutes an important part of the formal proof, as the combinatorial
arguments involved are notably lengthy and intricate at least when it comes to
formal proofs.

It is essential to note that ``monochromatic infinite outward path'' in the
context of the SSW theorem refers to walks where all vertices are distinct.  As
it is customary in graph theory, we will use the word walk when following
successive vertices moving in the direction of the arrows in a directed graph
and we will use the word path for a walk which never visits a vertex more than
once. While an infinite outward path in the asymmetric part of the transitive
closure of $\mathfrak{E}_{b}$ (resp.~for $\mathfrak{E}_{r}$)\footnote{note that for an asymmetric
  relation a walk is also a path.}  implies an infinite walk in
$\mathfrak{E}_{b}$ (resp.  in $\mathfrak{E}_{r}$), the latter may involve repeated vertices. A
challenge of this work lies in proving that an infinite outward path for
$\mathfrak{E}_{b}$ (resp.  in $\mathfrak{E}_{r}$) can be constructed.  We prove in
Theorem~\ref{th:Sands-Sauer-Woodrow-ext} the stronger version of the SSW theorem
and deduce the SSW theorem in Corollary~\ref{th:Sands-Sauer-Woodrow-orig}.

Directed graphs defined by symmetric relations $\mathfrak{E}_{r}$ and $\mathfrak{E}_{b}$ and containing
at least one outward path satisfy the assumptions of Theorem~\ref{th:Sands-Sauer-Woodrow-ext}
while the SSW Theorem cannot be applied.
As a simple example, consider the following infinite graph with only one color
$\dgraph= (\NN,\rel{E}_b)$ where for all $n\in \NN$ we have $n \rel{E}_b(n+1)$ and
$(n+1) \rel{E}_bn$.  This graph contains a monochromatic (here blue) infinite
outward path given by the injective function $f: n\in \NN \mapsto n\in \NN$ as we have
$f(n) \rel{E}_b f(n+1)$ for all $n \in \NN$. Thus,
SSW theorem cannot be used while it is clear that the
independent set $\na{0}$ satisfies the conclusions of the theorem.
By contrast, the relation $\rel{E}_b$ is
symmetric and thus the asymmetric part of the transitive closure of the binary
relation $\rel{E}_b$ is empty and we can conclude by the stronger
Theorem~\ref{th:Sands-Sauer-Woodrow-ext} that there exists an independent set
satisfying the conclusions of the SSW theorem.

To our knowledge, this is he first attempt to give a formal proof of the SSW
theorem for infinite graphs. By contrast, several formal proofs of the
Gale-Shapley algorithm exist. The first one was in Coq and given
in~\cite{Hamid-Castleberry}, the second one in Isabelle/HOL with several
implementations of the algorithm is detailed in~\cite{Nipkow}, and the third one
in Lean is available at \verb!https://github.com/mmaaz-git!\verb!/stable-marriage-lean!.

In Section~\ref{sec:backgrounds}, we provide background on binary relations and
graphs as binary relations.  In
Section~\ref{Binary_relations_and_graphs_in_Rocq}, we describe the Rocq
formalization that we use for binary relations and graphs using classical sets
as developed in MathComp, and we introduce some specific vocabulary and objects
needed for the SSW theorem, such as the asymmetric part of a relation,
independent sets, the monochromatic relation, and infinite walks/paths.  In
Section 4, we state a strengthened SSW theorem assuming no infinite paths in the
asymmetric parts of \((E_b^+)^{as}\) and \((E_r^+)^{as}\) and the key
Lemma~\ref{lem:infinite_path_implies} to relate the strengthened SSW theorem to
the original one.  In Section~\ref{rocq_proof_theorem}, we give the proof of the
strengthened SSW theorem (Theorem~\ref{th:Sands-Sauer-Woodrow-ext}) and in
Section~\ref{rocq_proof_coro}, we give a sketch of proof of
Lemma~\ref{lem:infinite_path_implies}.  Finally, in Section~\ref{github}, we
describe the \texttt{github} repository of Rocq files used for the formalization.

\section{Backgrounds}
\label{sec:backgrounds}

In \S\ref{sub:binary-rel}, we give backgrounds on binary relations
and in \S\ref{sub:graphs}, we provide background on graphs represented using
binary relations.

\subsection{Background on binary relations}
\label{sub:binary-rel}

Let $\AGENT$ be a nonempty set (finite or not).  We recall that a \emph{(binary)
  relation}~$\relation$ on~$\AGENT$ is a subset
$\relation \subset \AGENT\times\AGENT $ and that
\( \bgent\, \relation\, \cgent \) means \( \np{\bgent,\cgent} \in \relation \).
For any subset \( \Bgent \subset \AGENT \), the \emph{(sub)diagonal relation} is
\( \Delta_{\Bgent} = \bset{ \np{\bgent,\cgent} \in \AGENT\times\AGENT }%
{ \bgent=\cgent \in \Bgent } \) and the \emph{diagonal relation} is
\( \Delta=\Delta_{\AGENT} \). A relation is \emph{reflexive} if \(\Delta \subset \relation\).
A \emph{foreset} of a relation~$\relation$ is any set of the form
\( \relation \, \cgent = \defset{ \bgent \in \AGENT }{ \bgent\, \relation \,
  \cgent } \), where \( \cgent \in \AGENT \), or, by extension, of the form
\( \relation \, \Cgent = \defset{ \bgent \in \AGENT }{ \exists \cgent \in \Cgent \eqsepv
  \bgent\, \relation \, \cgent } \), where \( \Cgent \subset \AGENT \).
An \emph{afterset} of a relation~$\relation$ is
any set of the form \( \bgent \, \relation = 
\defset{ \cgent \in  \AGENT }{ \bgent\, \relation \, \cgent } \),
where \( \bgent \in \AGENT \), 
or, by extension, of the form \( \Bgent \, \relation = 
\defset{ \cgent \in  \AGENT }{ \exists \bgent \in \Bgent \eqsepv \bgent\,
  \relation \, \cgent } \), where \( \Bgent \subset \AGENT \).
The \emph{opposite} or \emph{complementary~$\Complementary{\relation}$} of a binary
relation~$\relation$ is the relation~$\Complementary{\relation}=\AGENT\times\AGENT\setminus\relation$,
that is, defined by \( \bgent\, \relation^{\mathsf{c}} \, \cgent \iff 
\neg \np{ \bgent\, \relation \, \cgent } \).
The \emph{converse~$\Converse{\relation}$} of a binary relation~$\relation$ is
defined by \( \bgent\, \Converse{\relation} \, \cgent \iff \cgent\, \relation \, \bgent
\) (and $\relation$ is  \emph{symmetric} if \( \Converse{\relation}=\relation \)).
The \emph{composition}
$\relation\relation'$ of two
binary relations~$\relation, \relation'$ on~$\AGENT$ is defined by
\( \bgent (\relation\relation') \cgent \iff
\exists \delta \in  \AGENT \), \( \bgent\, \relation \, \delta \)
and \( \delta\, \relation' \, \cgent \);
then, by induction we define\footnote{%
  In what follows, when we consider a binary relation
  as a subset $\relation \subset \AGENT\times\AGENT $, we will use the notation
  \( \SetProd{\relation}{n} \subset \SetProd{\AGENT\times\AGENT}{n}\), where $n$ is
  a positive integer, to
  denote a product subset of the product set~\( \AGENT^{2n} \),
  thus making the distinction with the binary relation
  $\relation^{n} \subset \AGENT\times\AGENT $ obtained by $n$~compositions.}
\( \relation^{n+1}=\relation\relation^{n} \) for \( n \in \NN^* \). 
The \emph{transitive closure} of a binary relation~$\relation$ is
\( \TransitiveClosure{\relation} = \cup_{k=1}^{\infty} \relation^{k} \)
(and $\relation$ is  \emph{transitive} if \( \TransitiveClosure{\relation}=\relation \))
and the \emph{reflexive and transitive closure} is 
\( \TransitiveReflexiveClosure{\relation}= \TransitiveClosure{\relation} \cup
\Delta = \cup_{k=0}^{\infty} \relation^{k} \) with the convention $\relation^0=\Delta$. 
A \emph{partial equivalence relation} is a symmetric and transitive binary
relation (generally denoted by~$\sim$ or~$\equiv$).
An \emph{equivalence relation} is a reflexive, symmetric and transitive binary
relation.

\subsection{Background on graphs as binary relations}
\label{sub:graphs}

Let $\VERTEX$ be a nonempty set (finite or not), whose elements are called
\emph{vertices}.  Let \( \EDGE \subset \VERTEX\times\VERTEX \) be a relation
on~$\VERTEX$, whose elements are ordered pairs (that is, couples) of vertices
called \emph{edges}.  The first element of an edge is the \emph{tail of the
  edge}, whereas the second one is the \emph{head of the edge}.  Both tail and
head are called \emph{endpoints} of the edge, and we say that the edge connects
its endpoints.  We define a \emph{loop} 
as an element of \( \Delta \cap \EDGE \), that is, a loop is an edge that connects a
vertex to itself.

A \emph{graph}, as we use it throughout this paper, is a
couple~$(\VERTEX,\EDGE)$.  This definition is basic, and we now stress
proximities and differences with classic notions in graph theory.  As we define
a graph, it may hold a finite or infinite number of vertices; there is at most
one edge that has a couple of ordered vertices as single endpoints, hence a
graph (in our sense) is not a multigraph (in graph theory); loops are not
excluded (since we do not impose $\Delta \cap \EDGE=\emptyset$).  Hence, what we call a graph
would be called a directed simple graph permitting loops in graph theory.

In the graph~$(\VERTEX,\EDGE)$, the \emph{directed edges} are the elements of
$\EDGE \cap \npComplementary{ \Converse{\EDGE} }$ --- that is, edges with
$(\cgent,\bgent)\in \EDGE$ such that $(\bgent, \cgent)\not\in \EDGE$ (recall that we
do not assume that $\EDGE \cap \Converse{\EDGE}=\emptyset$).  Then, the
graph~$(\VERTEX,\EDGE)$ is said to be \emph{directed} if all edges are directed
edges, or, equivalently, if \( \EDGE \cap \Converse{\EDGE}=\emptyset \), that is, when no
two edges have the same endpoints.  \medskip

\section{Binary relations and graphs in Rocq}
\label{Binary_relations_and_graphs_in_Rocq}

We formalize in this section binary relations (in \S\ref{sub:rel}), graphs and infinite paths
(in \S\ref{sub:graphs}) following the Rocq mathcomp library~\cite{MathComp:2022}.

\subsection{Binary relations as classical sets in Rocq}
\label{sub:rel}
We have chosen to implement binary relations as sets on a product space using the classical sets implemented in
\texttt{classical\_sets.v} from the Rocq mathcomp library~\cite{MathComp:2022}
using SSReflect tactics~\cite{Gonthier-Mahboubi-Tassi:2016}. 
\begin{alectryon}
  {\small \Snippetrelation{.\footnote{At the end of a Coq statement, ended by a
        dot belonging to the Vernacular (the language of Coq commands)
        we add a dot or a comma which serve as text punctuation.
        All the snippets of code are generated by Alectryon~\cite{Pit-Claudel}
        which ensures that code and documentation are in sync.}}}
\end{alectryon}
As described in more details below, we have developed a large library for
relations taking into account all the definitions recalled in \S\ref{sub:binary-rel} 
and we give in Table~\ref{tab:relation-rocq} the link between the
mathematical notations and the Rocq implementation.
\begin{table}[h]
  \centering
  \begin{tabular}{|c||c|c||c|c|}
    \hline
    name 
    & Notation
    & Rocq notation
    \\ \hline
    & relation $\relation$ on $X$
    & \verb!(R: relation T)!
    \\
    & $\bgent, \cgent \in X$
    & \verb!(! $\bgent$ $\cgent$\verb!: T)!
    \\
    & $\Bgent, \Cgent \subset X$
    & \verb!(! $\Bgent$ $\Cgent$\verb!:set T)!
    \\
    \hline
    relation
    & $\bgent \relation \cgent $
    & \verb! R (!$\bgent$,$\cgent$\verb!)!
      or \verb!(!$\bgent$\verb!,! $\cgent$\verb!) \in R!
    \\
    union
    & $\relation \cup \relation'$
    &\verb! R `|` R'!
    \\
    intersection 
    & $\relation \cap \relation'$
    & \verb!R `&` R'!
    \\
    subset
    &$\relation \subset \relation'$
    & \verb!R `<=` R'!
    \\
    opposite 
    &$\Complementary{\relation}$
    &\verb!R.^c!
    \\\hline
    diagonal relation
    & \( \Delta_{\Bgent} \) and  \( \Delta \)
    & $\Delta$\verb!_(!$\Bgent$\verb!)! and  $'\Delta$
    \\
    foreset
    &\( \relation \, \cgent \) and \( \relation \,\Cgent \)
    & \verb!R#_(!$\cgent$\verb!)! and \verb!R#!$\Cgent$
    \\
    afterset
    &\( \bgent \, \relation \) and \( \Bgent \, \relation \)
    & $\bgent$\verb!_:#R! and  $\Bgent$\verb!:#R!
    \\
    converse
    & $\Converse{\relation}$
    & \verb!R^-1!
    \\
    composition
    & $\relation\relation'$
    & \verb!R `;` R'! 
    \\
    $n$-composition
    &$\relation^{n}$
    & \verb!R^(n)!
    \\
    transitive closure
    & \( \TransitiveClosure{\relation}\)
    & \verb!R.+!
    \\
    transitive reflexive closure
    &$\TransitiveReflexiveClosure{\relation}$
    &\verb!R.*!
    \\ \hline
  \end{tabular}
  \caption{Correspondence table \label{tab:relation-rocq}}
\end{table}

For the proof of the SSW Theorem we also need some specific developments
namely defining the asymmetric part of a relation and defining the notion of $\relation$-independent sets.

\begin{definition}[Asymmetric part of a relation]
  For all $\relation$, $ x \Asym{\relation} y \iff x \relation y \wedge \neg\bp{x \Converse{\relation} y}$.
\end{definition}

\begin{alectryon}
  {\small \Snippetasym}
\end{alectryon}

\begin{definition}[$\relation$-independent sets]
  \label{def:independent-sets}
  Let $\relation$ be a relation on the set $\PRIMAL$, a subset $\Primal \subset \PRIMAL$ is said to be
  $\relation$-independent if
  \begin{equation}
    \forall (x,y) \in \Primal^2,  (x =y) \vee \neg (x \relation y).
  \end{equation}
  We denote by ${\cal I}_{\relation}$ the set composed of the $\relation$-independent subsets of $\PRIMAL$.
\end{definition}

\begin{alectryon}
  {\small \SnippetRelIndep}
\end{alectryon}
\subsection{Directed graphs and infinite paths in Rocq}
\label{sub:graphs}

A graph $(\VERTEX,\EDGE)$ is given in our Rocq implementation by an oriented
pair composed of a type \texttt{(T: Type)} and a relation on \texttt{T}, that is
$(\EDGE:\texttt{relation T})$ (which is equivalent to a set declaration
$(\EDGE:\texttt{set T*T})$).  Thus, the (classical) set definition of mathcomp
analysis \texttt{classical\_sets.v} is used to formalize a graph.  We have not
used the Coq package \texttt{graph-theory}~\cite{Doczkal-Pous-2029} to formalize
graph as we wanted to be able to manipulate infinite graphs.  To formalize a
directed graph whose edges are colored with two colors we consider two relations
defined on the same type \texttt{T} that is $\rel{E}_b$ and $\rel{E}_r$ giving
respectively the blue and the red edges.

Moreover, we introduce the monochromatic relation $\rel{M}$ given by the union
of the two transitive closures $\rel{E}_b^{+}$ and $\rel{E}_r^{+}$. That is two
vertices are related through the relation $\rel{M}$ if there exists a
monochromatic path between them.

\begin{definition}[Monochromatic relation $\rel{M}$]
  We consider two relations $\rel{E}_b$ and $\rel{E}_r$ together with their
  transitive closure $\rel{E}_b^{+}$ and $\rel{E}_r^{+}$ to define the
  monochromatic relation $\rel{M}$ defined by
  \begin{equation}
    \rel{M} = \rel{E}_b^{+} \cup \rel{E}_r^{+}
    \eqfinp
  \end{equation}
\end{definition}
\begin{alectryon}
  {\small \SnippetMono}
\end{alectryon}
We end this paragraph with Rocq definition for infinite (outward) walks and
infinite (outward) paths.

\begin{definition}[Infinite walk/path] A relation, $\relation$, on the set $\PRIMAL$
  is said to have an \emph{infinite (outward) walk} if there exists a mapping
  $f: \NN \to \PRIMAL$, such that
  \begin{equation}
    \forall n \in \NN\eqsepv f(n) \relation f(n+1)
    \eqfinp
  \end{equation}
  Moreover, if the function $f$ is injective, the relation $\relation$ is said
  to have an \emph{infinite (outward) path}.
\end{definition}

\begin{alectryon}
  {\small \Snippetiic}\\
  {\small \Snippetiicinj}
\end{alectryon}

\section{Rocq formulation of the SSW Theorem}
\label{sec:rocq-formulation}

We now formulate our main result (Theorem~\ref{th:Sands-Sauer-Woodrow-ext})
which is a strongest version of the SSW theorem and derive the SSW original
theorem (Corollary~\ref{th:Sands-Sauer-Woodrow-orig}) using the technical
Lemma~\ref{lem:infinite_path_implies}.
Note that in Theorem~\ref{th:Sands-Sauer-Woodrow-ext}
and Corollary~\ref{th:Sands-Sauer-Woodrow-orig} we have replaced the sentence ``there is a
monochromatic path from $x$ to a vertex of $S$'' by the statement
$x\in \rel{M} S$, that is \texttt{\PY{n}{x}~\(\in\)~\PY{n}{Mono}\PY{o}{\PYZsh{}}\PY{n}{Sm}} in Rocq.
The fact that the statement \texttt{\PY{n}{x}~\(\in\)~\PY{n}{Mono}\PY{o}{\PYZsh{}}\PY{n}{Sm}}
implies the existence of a monochromatic path from \texttt{x} to a vertex of the
set \texttt{Sm} is proved in the following Lemma~\texttt{\PY{n+nf}{Mono2path}}.
\begin{alectryon}
  {\small \SnippetMonotopath}.
\end{alectryon}

\begin{theorem} Let $\dgraph$ be a directed graph whose edges are colored with two
  colors and defined by two binary relations $\rel{E}_b$ and $\rel{E}_r$.
  Assume that the set of vertices is nonempty and that the
  relations $\Asym{\np{\rel{E}_b^+}}$ and $\Asym{\np{\rel{E}_r^+}}$ have no
  infinite walks.\footnote{Note that for asymmetric transitive relations walks are always
    paths.}  Then, there is an $\rel{M}$-independent set $S$ of vertices of $\dgraph$
  such that, for every vertex $x$ not in $S$, there is a monochromatic path from
  $x$ to a vertex of $S$.
  \label{th:Sands-Sauer-Woodrow-ext}
\end{theorem}
\begin{alectryon}
  {\small \SnippetMainTh}
\end{alectryon}

Finally, we consider the following lemma on infinite paths.
\begin{lemma} Let $\rel{R}$ be a relation. If there exists an infinite path for
  the relation $\Asym{\np{\rel{R}^+}}$, then there exists
  an infinite path for the relation~$\rel{R}$.
  \label{lem:infinite_path_implies}
\end{lemma}
\begin{alectryon}
  {\small \Snippetinfasym}.
\end{alectryon}
Using Lemma~\ref{lem:infinite_path_implies}, we obtain the original SSW Theorem as a corollary of
Theorem~\ref{th:Sands-Sauer-Woodrow-ext}.

\begin{corollary}[B. Sands, N. Sauer and R. Woodrow~\cite{Sands-Sauer-Woodrow-1982}]
  Let $\dgraph$ be a directed graph whose edges are colored with two colors and
  defined by two binary relations $\rel{E}_b$ and $\rel{E}_r$.  Assume that the
  set of vertices is nonempty and that the two relations
  $\rel{E}_b$ and $\rel{E}_r$ have no infinite paths.  Then, there is an
  $\rel{M}$-independent set $S$ of vertices of $\dgraph$ such that, for every vertex
  $x$ not in $S$, there is a monochromatic path from $x$ to a vertex of $S$.
  \label{th:Sands-Sauer-Woodrow-orig}
\end{corollary}
\begin{alectryon}
  {\small \SnippetSSWTh}
\end{alectryon}

\section{Rocq proof of Theorem~\ref{th:Sands-Sauer-Woodrow-ext}}
\label{rocq_proof_theorem}

The Rocq proof of Theorem~\ref{th:Sands-Sauer-Woodrow-ext} is given in this section
under the three assumptions: $(A_1)$
the set $\PRIMAL$ is nonempty; $(A_2)$ there does not exist an infinite
outward path for the binary relation~$\Asym{\np{\rel{E}_b^+}}$ and $(A_3)$ there
does not exist an infinite outward path for the binary
relation~$\Asym{\np{\rel{E}_r^+}}$.

We start by giving a sketch of proof and then the steps are detailed in subsections.
\begin{enumerate}
\item (In~\S\ref{subseq:poset}) \label{item:un} To any binary relation $\relation$ on the set $\PRIMAL$ we associate a
  relation (an order $\ROrder$) on the power set of $\PRIMAL$ using
  Equation~\eqref{ROrder}.  We prove in Lemma~\ref{lem:poset} that, when the
  relation $\relation$ is a sporder\footnote{A relation $\relation$ is a sporder if it is irreflexive and transitive.},
  then the pair
  $({\cal I}_{\relation}, \ROrder)$, is a partially ordered set, where
  ${\cal I}_{\relation}$ is the set of $\relation$-independent subsets of
  $\PRIMAL$ as defined in Definition~\ref{def:independent-sets}.
\item (In~\S\ref{subseq:scal}) We specialize step~\ref{item:un} to the sporder $\AstOrder$. We define a
  subset ${\cal S}$ of ${\cal I}_{\rel{M}}$ in Definition~\ref{def:Scal}.  We
  show that ${\cal S}$ is nonempty and using the fact that
  ${\cal S} \subset {\cal I}_{\rel{M}} \subset {\cal I}_{{\Asym{\np{\rel{E}_b^+}}}}$ we obtain that
  $({\cal S}, \AstOrder)$ is also a partially ordered set. We use $\Order$ as a
  simplified notation for $\AstOrder$.
  
\item (In~\S\ref{subseq:chains})\label{step:maximal}
  We consider chains $\mathcal{C}$ in $({\cal S}, \Order)$. To each chain $\mathcal{C} \subset {\cal S}$ we associate
  a subset $S^{\infty}_{\mathcal{C}}$ of $\PRIMAL$ as defined in Definition~\ref{def:Scal_chain_inf}.
  We prove in Lemma~\ref{lem:Sinfisupper} that for all  $S \in \mathcal{C}$, we have that $S \Order S^{\infty}_{\mathcal{C}}$ and
  $S^{\infty}_{\mathcal{C}}$ belongs to ${\cal S}$.
  We conclude by Zorn's lemma that there exists a maximal set in ${\cal S}$.
\item (In~\S\ref{subseq:maximal}) The last step consists in proving that a maximal set in ${\cal S}$ (whose
  existence was proved in step~\ref{step:maximal}) is precisely the set which
  satisfies the conclusions of Theorem~\ref{th:Sands-Sauer-Woodrow-ext}.
\end{enumerate}

\subsection{A partially ordered set $({\cal I}_{\relation}, \ROrder)$}
\label{subseq:poset}
\begin{definition}[An order relation on sets induced by a relation]
  \label{def:set_relation}
  Let $\relation$ be a relation on the set $\PRIMAL$.
  For any subsets  $\Primal, \Dual \subset \PRIMAL$, we define an order $\ROrder$ defined by
  \begin{equation}
    \Primal \ROrder \Dual \iff
    \forall \primal \in \Primal \eqsepv \exists \dual \in \Dual \text{ s.t. } (\primal =\dual) \vee \bp{\primal \relation \dual}
    \eqfinv
    \label{ROrder}
  \end{equation}
  or in a more compact form using notations from Table~\ref{tab:relation-rocq}
  \begin{equation}
    \Primal \ROrder \Dual \iff \Primal \subset \Dual \cup \relation \Dual
    \eqfinp
  \end{equation}
  
\end{definition}

\begin{alectryon}
  {\small \Snippetleset}.
\end{alectryon}

We have that
\begin{equation}
  \forall \Primal \eqsepv \Primalbis \subset \PRIMAL \eqsepv
  \Primal \subset \Primalbis \implies \Primal \ROrder \Primalbis
  \eqfinp
\end{equation}

\begin{alectryon}{\small \SnippetlesetI}\end{alectryon}

\begin{lemma}
  \label{lem:poset}
  When the relation $\relation$ is a sporder, then the set
  $({\cal I}_{\relation}, \ROrder)$ is a partially ordered set.
  \footnote{The
  relation $\ROrder$ is reflexive, antisymmetric and transitive on
  ${\cal I}_{\relation}$ the $\relation$-independent subsets of $\PRIMAL$.}
  {\normalfont 
    \begin{alectryon}{\small \Snippetlesetporder}\end{alectryon}
    }
\end{lemma}

\subsection{A subset ${\cal S} \subset {\cal I}_{\rel{M}}$ and the partial order $({\cal S},\AstOrder)$}
\label{subseq:scal}

We consider here a subset $\mathcal{S} \subset {\cal I}_{\rel{M}}$ of the collection of $\rel{M}$-independent sets defined
as follows. A subset $S \subset \PRIMAL$ belongs to $\mathcal{S}$ if it is a
$\rel{M}$-independent nonempty set and for all $v \in \PRIMAL$, if there exists a red
path from $S$ to $v$ then there exists a monochromatic path from $v$ to
$S$.
\begin{definition}
  \label{def:Scal}
  We define ${\cal S}$ as the set of the $\rel{M}$-independent nonempty subsets
  of $\PRIMAL$ which satisfy the following property. The afterset of $S$ for the
  $\rel{E}_r^{+}$ relation is included in the foreset of $S$ for the $\rel{M}$
  relation.  More formally,
  \begin{equation}
    {\cal S} =
    \bset{ S \subset \PRIMAL}{ S \text{ is $\rel{M}$-independent}
      \eqsepv
      S \rel{E}_r^{+} \subset \rel{M} S
      \eqsepv
      S \not= \emptyset}
    \eqfinp
  \end{equation}
\end{definition}
\begin{alectryon}
  {\small \SnippetScal}
\end{alectryon}
As a first property we have that $\mathcal{S}$ is nonempty.
\begin{lemma}
  Note that the set $\mathcal{S}$ is nonempty by Assumption~$(A_2)$. We can prove that 
  there is a vertex $v$ such that $v {\rel{E}_r^{+}} y$ implies
  $y {\rel{E}_r^{+}} v$ for all $y$, whence $\{v\} \in \mathcal{S}$, since otherwise we could construct an 
  infinite outward path for $\Asym{\np{\rel{E}_r^+}}$ giving a contradiction with~$(A_2)$.
\end{lemma}
\begin{alectryon}
  {\small \SnippetScalnotempty}
\end{alectryon}

We need now to stick to sets contained in ${\cal S}$ and equip the set ${\cal S}$ with the relation $\AstOrder$. 
For that purpose, we define in Rocq a new sigma-type which is precisely the type of sets contained in ${\cal S}$, 
\begin{alectryon}
  {\small \SnippetSType}
\end{alectryon}
and denote by $\Order$, the restriction of the relation $\AstOrder$ to the set ${\cal S}$. 
The Rocq notation we use for the restricted relation id \texttt{\PY{o}{[\PYZlt{}=]}} defined by.
\begin{alectryon}
  {\small \SnippetleSetone}
\end{alectryon}

Finally, we prove that $({\cal S},  \leqslant_{\rel{E}_b^{+}})$ is 
also a partially ordered set using arguments similar to the one used in Lemma~\ref{lem:poset}.

\subsection{Chains in $({\cal S}, \Order)$ and upper bound candidate $S^{\infty}_{\mathcal{C}}$}
\label{subseq:chains}
Now, to any subset ${\mathcal{C}} \subset {\cal S}$ we associate a new set $S^{\infty}_{\mathcal{C}}$ defined as follows.
\begin{definition}[$S^{\infty}_{\mathcal{C}}$]
  \label{def:Scal_chain_inf}
  Let $\mathcal{C} \subset {\cal S} $ be given, the subset of vertices of $\PRIMAL$ that belong to every member of
  $\mathcal{C}$ from some point on is denoted by $S^{\infty}_{\mathcal{C}}$. More precisely $S^{\infty}_{\mathcal{C}}$ is defined by
  \begin{equation}
    S^{\infty}_{\mathcal{C}}=\Bset{s \in \PRIMAL}{\exists S \in \mathcal{C} \text { s.t. } s \in U
      \text { whenever } U \in \mathcal{C} \text { and } S \Order U}
  \end{equation}
\end{definition}
\begin{alectryon}{\small \SnippetSinf}\end{alectryon}
We consider chains of sets in ${\cal S}$ defined as follows
\begin{alectryon}
  {\small \Snippetchains}\\
  {\small \SnippetChains}
\end{alectryon}

\begin{lemma} Let $\mathcal{C} \subset {\cal S} $ be given and nonempty.
  Under Assumptions~$(A_1)$ and~$(A_2)$, the set $S^{\infty}_{\mathcal{C}}$
  is nonempty and such that for all  $S \in \mathcal{C}$, we have that $S \Order S^{\infty}_{\mathcal{C}}$.
  Moreover, when $\mathcal{C}$ is a $(\Order)$-chain in {\cal S}, we have that $S^{\infty}_{\mathcal{C}}$ belongs to ${\cal S}$.
  \label{lem:Sinfisupper}
\end{lemma}

The proof in the Rocq prover follows the following lines.
Let $\mathcal{C} \subset {\cal S} $ be given. We consider the subset of $\PRIMAL$ given by
$\cup_{S \in {\cal C}} S$ and define a relation $\rel{C}$ of this set (denoted \texttt{\PY{n+nf}{RC}} in Rocq).

\begin{equation}
  \rel{C} = \Delta_{S^{\infty}} \cup  \Delta_{\Complementary{\np{S^{\infty}}}} \Asym{\bp{\rel{E}_b^+}}
  \label{def:relC}
\end{equation}

\begin{alectryon}{\small \SnippetElt}\end{alectryon}
\begin{alectryon}{\small \SnippetRC}\end{alectryon}

Using the definition of $S^{\infty}$, we prove that the relation $\rel{C}$ as defined in 
Equation~\eqref{def:relC} is (left) total 
\begin{definition}[Total] A relation, $\relation$, on the set $\PRIMAL$ is said to be
  \emph{total} (or more precisely \emph{left total})
  if for all $\primal \in \PRIMAL$ the afterset $\na{x} \relation \not= \emptyset$.
\end{definition}

\begin{alectryon}
  {\small \Snippetlefttotal}\\
  {\small \SnippettotalRC}
\end{alectryon}

Using axiom of choice we can obtain for any $s\in\cup_{S \in {\cal C}} S$ a mapping
such that $s=f(0)$ and for all $n \in \NN$, $f(n) \rel{R} f(n+1)$. Then, we prove that 
there must exist an element of the image of function $f$ in the set $S^{\infty}$; otherwise 
we would obtain a contradiction with Assumption~$(A_3)$. 
\begin{alectryon}{\small \SnippettotalRCPTr}\end{alectryon}

It follows that set $S^{\infty}_{\mathcal{C}}$ is not empty and that for all  $S \in \mathcal{C}$, we have that $S \leqslant S^{\infty}_{\mathcal{C}}$.
\begin{alectryon}{\small \SnippetChooseRCCi}\end{alectryon}
\begin{alectryon}{\small \SnippetChooseRCSi}\end{alectryon}

\begin{alectryon}{\small \SnippetSinfScalP}\end{alectryon}

\begin{alectryon}{\small \SnippetSinfScal}\end{alectryon}

Finally, we prove that the set  $S^{\infty}_{\mathcal{C}}$ belongs to $\mathcal{S}$ 
and we obtain a maximal set in ${\cal S}$ using  Zorn Lemma. 


\begin{alectryon}{\small \SnippetIsMaximal}\\
  {\small \SnippetSmax}
\end{alectryon}

\subsection{The maximal element of ${\cal S}$ satisfies the conclusion of Theorem~\ref{th:Sands-Sauer-Woodrow-ext}}
\label{subseq:maximal}

The last step of the proof consists in proving that the maximal element of
${\cal S}$ satisfies the conclusion of Theorem~\ref{th:Sands-Sauer-Woodrow-ext}.
Thus considering a set $S_m$ (\texttt{\PY{n}{Sm}} in Rocq) we want to prove that when $S_m$ is maximal
(that is it satisfies \texttt{\PY{n}{IsMaximal}~\PY{n}{Sm}}), then the set $S_e$ (\texttt{\PY{n}{Se}} in Rocq) is empty, where
\texttt{\PY{n}{Se}} is defined by
\begin{alectryon}
  {\small \SnippetSx}.
\end{alectryon}
The proof is by contradiction. We assume that $S_e \not= \emptyset$ and prove that it leads to a contradiction.
As a preliminary fact, if we assume that the set $S_e$ is nonempty then there
must exist an element $x\in S_e$ satisfying the additional property that for all
$y\in S_e$, if $x \rel{E}_r^+ y$ then $y \rel{E}_r^+ x$, as otherwise we could
contradict Assumption~$(A_2)$.
\begin{alectryon}{\small \SnippetSxm} \\
  {\small \SnippetSxone}
\end{alectryon}
Now, to each element $x$ we associate the set $T_{m,x} \subset S_m$
(\texttt{\PY{n}{Tm}~\PY{n}{x}} in Rocq) defined by
\begin{alectryon}{\small \SnippetTm}, \end{alectryon}
satisfying the two following lemmas.
\begin{alectryon}{\small \SnippetTmI}  \\
  {\small \SnippetSbunp}.
\end{alectryon}

Now, following the logical steps established in the original
publication~\cite{Sands-Sauer-Woodrow-1982} we prove some lemmas that
finally give
\begin{alectryon}{\small \Snippetfactnine} \\
  {\small \Snippetfactten} \\
  {\small \Snippetfacteleven},
\end{alectryon}
from which Theorem~\ref{th:Sands-Sauer-Woodrow-ext} follows.

\section{Rocq proof of Corollary~\ref{th:Sands-Sauer-Woodrow-orig}}
\label{rocq_proof_coro}

As already explained in Sect.~\ref{sec:rocq-formulation}, the proof of
Corollary~\ref{th:Sands-Sauer-Woodrow-orig} is a simple consequence of
Lemma~\ref{lem:infinite_path_implies}, which we restate here. If there exists an
infinite path for the relation $\Asym{\np{\rel{R}^+}}$, then there exists an
infinite path for the relation~$\rel{R}$. We briefly describe now the steps used
in the Rocq proof of Lemma~\ref{lem:infinite_path_implies}. We formalize paths
(and walks) with sequences.  We prove in Lemma
\texttt{\PY{n+nf}{TCP\PYZus{}uniq}} that two vertices are related by the
transitive closure of a relation $\rel{R}$ if and only if there exists a
sequence of vertices giving a path for the relation $\rel{R}$ joining them.
\begin{alectryon}{\small \Snippettcpuniq}, \end{alectryon} where
\texttt{\PY{n}{allL}~\PY{n}{R}~\PY{n}{s}~\PY{n}{x}~\PY{n}{y}\PY{o}} means
that the successive elements of the sequence \texttt{x::(rcons s y)}
are related by the relation \texttt{R}.

Thus, if $x \Asym{\np{\rel{R}^+}}y$ we can build a $\mathtt{uniq}$ sequence
$(x_0, x_1,\ldots x_p, x_{p+1})$ with $x_0=x$ and $x_{p+1}=y$ where two consecutive elements are related
by the relation $\rel{R}$ as illustrated on the following picture.
\begin{center}
  \includegraphics[width=0.4\textwidth]{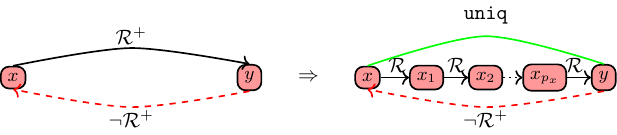}
\end{center}
    
Now, if we have $x \Asym{\np{\rel{R}^+}}y$ and $y \Asym{\np{\rel{R}^+}}z$ we can apply the previous result twice 
with two sequences  $(x_1, \ldots, x_p)$ and $(y_1,\ldots, y_q)$ as illustrated on the following picture.
\begin{center}
  \includegraphics[width=0.8\textwidth]{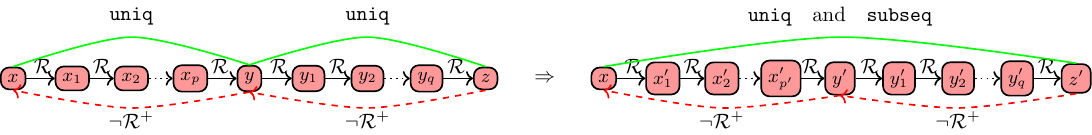}
\end{center}
However, we can have elements shared by the two sequences $(x_1, \ldots, x_p)$ and $(y_1,\ldots, y_q)$.
We show, in a key intermediate lemma \texttt{\PY{n+nf}{Asym2P5'}},
that we can find a subsequence of the sequence  $(x_1, \ldots, x_p)$
denoted by  $(x'_1, \ldots, x'_p)$ and a new vertex $y'$, and a sequence $(y'_1,\ldots, y'_q)$ in such a way that the whole sequence
$(x,x'_1, \ldots, x'_{p'}$, $y', y'_1,\ldots $, $y'_{q'},z)$ is a $\mathtt{uniq}$ sequence
while preserving the fact that we have $x \Asym{\np{\rel{R}^+}}y'$ and $y' \Asym{\np{\rel{R}^+}}z$.
In this step it is important that $(x'_1, \ldots, x'_{p'})$ is a subsequence of $(x_1, \ldots, x_p)$ to enable a recursive
construction.

Using the intermediate lemma \texttt{\PY{n+nf}{Asym2P5}} combined with the Axiom of dependent choice we can build two mappings
\texttt{k} and \texttt{l} satisfying
\begin{alectryon}{\small \SnippetAsymdPV},\end{alectryon}

Now, to obtain the final result we have to combine \texttt{k} and \texttt{l} into a unique mapping $l$
visiting the successive elements of the walk described by  \texttt{k} and \texttt{l} and prove that
the function $l$ is injective. Using Lemma \texttt{\PY{n+nf}{Asym2P5}}, we have that two consecutive sequences \texttt{(l n)} and
\texttt{(l n.+1)} do not share elements, and then we use the remaining asymmetry between consecutive elements described by the mapping
\texttt{k}, that is
\texttt{\PY{o}{\(\neg\)}~\PY{n}{R}\PY{o}{.+}~\PY{o}{(}\PY{n}{k}~\PY{n}{n}\PY{o}{.+}\PY{l+m+mi}{1}\PY{o}{,}~\PY{n}{k}~\PY{n}{n}\PY{o}{)}},
to prove that elements of sequences $\texttt{(l n)}$ and \texttt{(l n.+p)} cannot share elements if \texttt{p > 1}.
This leads to the Rocq version of Lemma~\ref{lem:infinite_path_implies}.

\section{\texttt{Github} companion repository}
\label{github}
The Coq code developed for proving the results exposed in this paper (and the results
exposed in two other papers related to causal graphs~\cite{Chancelier-De-Lara-Heymann-2024}
and~\cite{De-Lara-Chancelier-Heymann-2021})
is publicly available on GitHub\footnote{at URL
  \texttt{https://github.com/jpc-cermics/relations.git}}
and use Mathcomp/SSReflect~\cite{MathComp:2022,Gonthier-Mahboubi-Tassi:2016,Mahboubi-Tassi:2022}.
The specific files used for this paper are detailed in Table~\ref{tab:library}.

\begin{table}[h]
  \centering
  \begin{tabular}{|c||c|r|}
    \hline
    file name 
    & contents 
    & loc
    \\ \hline
    \texttt{rel.v} &  binary relations as sets  & 1516\\\hline
    \texttt{seq1.v} &  graph as binary relations and walks  & 1011 \\\hline
    \texttt{seq2.v} &  graph as binary relations and paths  & 394 \\\hline
    \texttt{paper\_monochromatic.v} & companion developments for this paper& 1121 \\\hline
    \texttt{paper\_monochromatic\_f.v} & proof of Lemma~\ref{lem:infinite_path_implies}& 951 \\\hline 
  \end{tabular}
  \caption{subset of \texttt{relations.git} repository \label{tab:library} used in this paper}
\end{table}

\section{Conclusion}

In this paper, we give a complete Rocq formalization of the Sands--Sauer--Woodrow
theorem following closely the original combinatorial argument. It appears that,
it was possible to strengthen the assumptions as given in
Theorem~\ref{th:Sands-Sauer-Woodrow-ext}, where the absence of infinite
monochromatic outward paths is replaced by the weaker requirement that the
asymmetric parts of the transitive closures of the red (resp.~the blue) relation
admit no infinite outward paths. Establishing the logical link back to the
classical formulation (Corollary~\ref{th:Sands-Sauer-Woodrow-orig}) relies on
the lemma showing how an infinite path in \(\Asym{(\rel{R}^+)}\) yields an
infinite path for \(\rel{R}\) itself, and it accounts for a substantial part of
the formal effort. This formal proof is accompanied by a Mathcomp library
devoted to binary relations (as sets) and paths/walk in graphs defined by binary
relations. This library was originally developed by the author for studying
causality on graphs as developed in~\cite{Chancelier-De-Lara-Heymann-2024}
and in~\cite{De-Lara-Chancelier-Heymann-2021}. This paper was also an attempt to
show that the library was reusable and extendable for other formalizations in
infinite graph theory and in areas where binary relations are primitive objects.
 
\newcommand{\noopsort}[1]{} \ifx\undefined\allcaps\def\allcaps#1{#1}\fi

\end{document}